Natural radioactive environments as sources of local disequilibrium for the emergence of life

Thiago Altair[1,5*†], Larissa M. Sartori[2], Fabio Rodrigues[3], Marcio G. B. de Avellar[4], Douglas Galante[1,5]

[1] Brazilian Synchrotron Light Laboratory (LNLS), Brazilian Center for Research in Energy and Materials (CNPEM). Av. Giuseppe Máximo Scolfaro, 10000, 13083-100, Campinas/SP, Brazil.

[2] Instituto de Matemática e Estatística, Universidade de São Paulo. Rua do Matão, 1010, 05508-090, São Paulo/SP, Brazil.

[3] Departamento de Química Fundamental Instituto de Química, Universidade de São Paulo. Av. Prof. Lineu Prestes, 748, 05508-000, São Paulo/SP, Brazil.

[4] Instituto de Astronomia, Geofísica e Ciências Atmosféricas, Universidade de São Paulo. Rua do Matão, 1226, 05508-090, São Paulo/SP, Brazil.

[5] Instituto de Física de São Carlos, Universidade de São Paulo. Av. Trabalhador São-carlense, 400, 13566-590, São Carlos/SP, Brazil.

[*] Author to whom correspondence should be addressed (thiago.altair.ferreira@usp.br)
[†] Present address: Instituto de Química de São Carlos, Universidade de São Paulo. Av. Trabalhador São-carlense, 400, 13566-590, São Carlos/SP, Brazil.

**Abstract**

Certain subterranean environments of Earth have naturally accumulated long-lived radionuclides, such as $^{238}$U, $^{232}$Th and $^{40}$K, near the presence of liquid water. In these natural radioactive environments (NRE), water radiolysis can produce chemical species of biological importance, such as $H_2$. Although the proposal of radioactive decay as an alternative source of energy for living systems has existed for more than thirty years, this hypothesis gained strength after the recent discovery of a peculiar ecosystem in a gold mine in South Africa, whose existence is dependent on chemical species produced by water radiolysis. In this work, we calculate the chemical disequilibrium generated locally by water radiolysis due gamma radiation and analyse the possible contribution of this disequilibrium for the emergence of life, considering conditions of early Earth and having as reference the alkaline hydrothermal vent (AHV) theory. Results from our kinetic model points out the similarities between the conditions caused by water radiolysis and those found on alkaline hydrothermal systems. Our model produces a steady increase of pH with time, which favours the precipitation of minerals with catalytic activity for protometabolism, as well as a natural electrochemical gradient in this aqueous environment. In conclusion, we described a possible free-energy conversion mechanism that could be a requisite for emergence of life in Hadean Earth.

1. **INTRODUCTION**

It is well known that the subsurface of Earth hosts most of prokaryotic biomass of the planet (Whitman *et al.*, 1998; Labonté *et al.*, 2015). At depths greater than 1 km below sea level, the environments have extreme conditions compared to superficial or oceanic ones, such as absence of oxygen, nutrients, sunlight and sparse sources of energy (Omar *et al.*, 2003; Colman *et al.*, 2017). In some of these deep regions, there are natural deposits of radioactive minerals, such as at gold mines of the southern part of Africa (Lippmann *et al.*, 2003; Omar *et al.*, 2003; Chivian *et al.*, 2008; Adam *et al.*, 2018), or at uranium deposits in Canada (Dubessy *et al.*, 1988; Richard *et al.*, 2012). Recently, it has been proposed that radioactive environments may be an interesting scenario for the emergence of life on early Earth, considering those on the surface (Adam, 2007) or in deep underground (Ebisuzaki and Maruyama, 2017) due to the chemical diversity produced by water radiolysis (see equation 1) and capacity to produce chemical energy comparable to that of the Urey-Miller experiment (Cataldo and Iglesias-Groth, 2017; Ebisuzaki and Maruyama, 2017). Also, as this work details, preliminary analysis showed that some of these natural radioactive environments (NRE) have similar physicochemical conditions when compared to alkaline hydrothermal vents (AHV) systems, now considered one of the most promising environments for the emergence of life (Russell *et al.*, 1989; Martin *et al.*, 2008; Lane and Martin, 2012; Sojo *et al.*, 2016; Ebisuzaki and Maruyama, 2017).

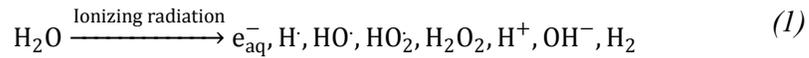

$$H_2O \xrightarrow{\text{Ionizing radiation}} e^-_{aq}, H^\cdot, HO^\cdot, HO_2^\cdot, H_2O_2, H^+, OH^-, H_2 \qquad (1)$$

It has already been postulated the importance of water radiolysis as a source of energy and of chemical species to sustain a living system some decades ago (Draganic and Draganic, 1971; Draganić *et al.*, 1983; Draganic *et al.*, 2005), but only after the work by Chivian et al. (2008) this hypothesis has shown to be factual. The authors described a lithoautotrophic microbial species whose survival is dependent on redox reactions involving radiolysis-produced compounds. The *Candidatus Desulforudis audaxviator* bacterium was found prevailing in a deep subsurface region around 2.8 km below the surface level, in fracture water at the Mponeng gold mine, located in the Witwatersrand basin, South Africa, in a hot (>60ºC) and alkaline (pH=9.3) environment (Lin *et al.*, 2006; Chivian *et al.*, 2008; Zhou *et al.*, 2011). Interestingly, the temperatures of the hydrothermal fluid formed in the vents of Lost City Hydrothermal Field (LCHF) system are in the range between 40ºC and 75ºC, and the pH, are in the range between 9 and 11. The difference between both environments are related to the energy source. The formation of hydrothermal fluid and also the energy source in these systems are the result of serpentinization reaction (Lowell, 2002; Russell and Arndt, 2004; Martin *et al.*, 2008) (see details in supplementary material).

Despite the importance of energy sources and disequilibrium for life as we know, it is the mechanism of free energy conversion (or entropy conversion) that maintains a living system in its necessary far-from-equilibrium condition (Schrodinger, 1946; Schneider and Kay, 1995; Russell *et al.*, 2013; Branscomb *et al.*, 2017). Modern organisms use complex and specific protein-based nanoengines driven by chemiosmotic gradients linked to metabolic pathways (Boyer, 1997; Branscomb and Russell, 2013; Sousa *et al.*, 2013; Sojo *et al.*, 2016), a biological universal setting (Goodsell, 2009; Branscomb and Russell, 2018). In these mechanisms, the chemiosmotic gradients are maintained by semipermeable membranes that restrict the diffusion of chemical species, while the crossing through these membranes is mediated by specific and also complex converters, leading to the transduction of the chemical energy stored in the gradients (Branscomb and Russell, 2013; Russell *et al.*, 2013). Finally, the AHV theory stands that hydrothermal systems that were present in early Earth could provide an analogous and also requisite setting to free-energy conversion for existence of living systems as we know (Russell and Arndt, 2004; Martin *et al.*, 2008; Lane *et al.*, 2010; Branscomb and Russell, 2013). Instead of modern complex enzymes, in AHV scenario, minerals would work as catalyzers coupling the generated chemiosmotic gradient to protometabolic pathways (Russell and Martin, 2004; Muñoz-Santiburcio and Marx, 2016; Muchowska *et al.*, 2017; Varma *et al.*, 2017). In this theory, the coupling would result in free-energy conversion and those mineral catalyzers may be constituted of iron and sulfur minerals (Fe-S minerals), or iron oxihydroxides, the so-called green rusts;

and driven by pH, electrochemical gradients between the hydrothermal fluid and early ocean (Branscomb and Russell, 2013; Ang *et al.*, 2015; Yamamoto *et al.*, 2017; Barge *et al.*, 2018; Ooka *et al.*, 2019). Thus, this could be a requisite for the emergence of life in Hadean Earth.

The Fe-S minerals such as greigite ($Fe_3S_4$) or mackinawite (FeS) are cited on works related to AHV theory to have been incorporated as clusters into the precipitates of the vents under the conditions of the early Earth. These propositions are based on early ocean composition and hydrothermal fluid conditions (Wächtershäuser, 1997; Russell and Martin, 2004; Sojo *et al.*, 2016). Also, an important feature is that the clusters of the Fe-S minerals are morphologically similar to active sites of important enzyme cofactors in the Acetyl-CoA synthesis pathway (Russell and Martin, 2004; Sojo *et al.*, 2016). This pathway unites carbon and energy metabolisms in Bacteria and Archaea domains, and it is expected to have been present in the Last Universal Common Ancestor (LUCA) (Lane *et al.*, 2010; Sojo *et al.*, 2016)(see supplementary material).

Thus, in this work, we have developed a numerical model to analyse the generation of chemical disequilibrium by water radiolysis due to gamma radiation from radioactive decay of $^{238}$U and $^{232}$Th-containing minerals, and $^{40}$K dissolved in early ocean waters. Besides, we estimated the influence of radiolysis on conditions found in habitable NRE, such as those that host *Ca. D. audaxviator*. Also, considering that similar physicochemical conditions and the predominance of reducing species are found in NRE, similarly to alkaline vents systems, we have analyzed a mineral-based transduction model for these systems based on AHV models. In summary, we aim to evaluate the possible contribution of NRE to the origin of life in the Hadean Earth (4Ga) based on analogous thermodynamic and kinetic arguments that stand the AHV theory. To this purpose, we have calculated how water radiolysis contributes to the aforementioned conditions in NRE. Finally, considering that NRE have no dependence on a particular mineral interaction as energy source, as occurs to alkaline hydrothermal vents, and its existence is a direct consequence of formation of rocky celestial bodies (Altair *et al.*, 2018), it was possible to extend this origin of life scenario to a widely present one, not restricted to early Earth (Bouquet *et al.*, 2017; Altair *et al.*, 2018). Other examples are the icy moons of the giant planets - such as Europa and Enceladus - which are the main targets of the next major missions for the search for signals of life on the Solar System (Grasset *et al.*, 2013).

## 2. Methods

### 2.1. Physicochemical conditions of the NRE and effects of water radiolysis in local chemical diversity

The geochemical analysis for radioactive environments in which *Ca. D. audaxviator* was discovered is the starting point for the numerical models developed here. We considered long-lived radionuclides with concentrations comparable to the ones found in the region of the geological mineralized and non-mineralized strata in Witwatersrand. Also, we used the concentration for $^{40}K$ that is estimated to be present at early oceans according to Draganić *et al*. (1991), also considering it as homogenously dissolved. As a control, we performed the calculations using carbonate chondrite long-lived radionuclide concentrations, since they are primitive bodies of the Solar System. Used in some works as model for early Earth (Javoy, 1995; Maruyama and Ebisuzaki, 2017). Numerical corrections was made for the decay rates to reproduce the conditions 4 Ga ago were performed for all radionuclides. Table 1 - Radionuclides used for the modelling of chemical species' production by water radiolysis shows the concentrations used in our model.

Table 1 - Radionuclides used for the modelling of chemical species' production by water radiolysis

| Radionuclide | Natural abundance (atom %)[a] | Concentration in chondrite (ppb)[b] | Concentration in mineralized strata in Witwatersrand (ppm)[c] | Concentration in non-mineralized strata in Witwatersrand (ppm)[c] | Half-life (years)[a] | γ decay energy (MeV/decay)[d] |
|---|---|---|---|---|---|---|
| $^{40}K$ | 0.0117(1) | 105 | 380* | 380* | $1.25 \times 10^9$ | 0.1566 |
| $^{232}$Th | 100 | 40 | 15 | 11 | 1.4x10$^{10}$ | 2.2447 |
| $^{238}$U | 99.2742(10) | 12 | 271 | 3 | 4.46x10$^{9}$ | 1.7034 |

Notes

[a]LIDE, 2003 (Lide, 2003)

[b]WAITE et al., 2017a (Waite *et al.*, 2017)

[c]LEFTICARIU et al., 2010 (Lefticariu *et al.*, 2010)

[d]BLAIR et al., 2007 (Blair *et al.*, 2007)

*Concentration estimated for early Earth (Draganić *et al.*, 1991)

For the models developed here, we considered the escape yield (G) constant, and its values are found in Pastina and Laverne (2001). Based on escape yields, we calculated the production rate of primary products as function of time using equations (2) and (3) (see Hoffmann (1992), Spinks, J.W.T.; Woods (1964), Blair *et al.* (2007) and Lin *et al.* (2005a), for example).

$$D_{eff} = \frac{\rho \cdot \sum_n D_n}{\frac{1}{1-\varphi} + \frac{1}{S \cdot \varphi}} \quad (2)$$

$$Y_P = \sum_P D_{eff} \cdot G_p \quad (3)$$

The two indices, $n$ and $p$, represent, respectively, the referred radionuclide ($^{238}$U, $^{232}$Th or $^{40}$K) and the radiolysis product; $Y_P$ is the rate of formation of the radiolysis product $p$ in mol. (L.s)$^{-1}$; $\rho$ represents the density of the local rock matrix in g.cm$^{-3}$; $D_n$, the dose of gamma radiation emitted by the radionuclide $n$ in MeV.(kg.s)$^{-1}$ as calculated in equation (3); $\varphi$ represents the rock porosity, which we considered to be equal to 0.1 (which is the actual porosity observed in Witwatersrand (Lin *et al.*, 2005a) and the maximum value found in depths of 1000 m in the Earth's crust (Vance *et al.*, 2007)); and we represent by $S$ the stopping power of the rock matrix (as in Lin (2005) and Blair (2007), with a value S = 1.14 for gamma radiation used here). Finally, the value of G is in mol.MeV$^{-1}$.

$$D_n = \frac{E \cdot \lambda \cdot c \cdot N_A}{A_n} \quad (4)$$

In equation (4), $E$ [MeV / decay] is the decay energy corrected for the loss of neutrinos via beta decay for 4Ga radionuclides; $\lambda = 1 / T_{1/2}$ [decay per year] is the decay constant, c [ppm] is the radionuclide concentration, $N_A$ is the Avogadro constant and $A_n$ [g / mol] is the radionuclide molar mass.

To evaluate the effects of long-term radiolysis, we used the model based on the kinetic one presented in Pastina and La Verne (2001) (Pastina and Laverne, 2001), testing different set of equations that resulted in stable numerical results. The effect aforementioned results in a chemical diverse water solution referred as "radiolytic fluid" throughout this paper. The chemical reactions used in the model are shown in Table S1. From that system of reactions, it was developed a system of ordinary differential equations (ODEs) based on rate law for each chemical equation. The chemical reactions show different orders of magnitude for chemical reaction rates. As a result, we ended with a stiff ODE system, which requires the application of appropriated methods for integration in order to get convergent and stable numerical solutions. Because of this, we used a Rosenbrock method for calculations. This is a class of implicit single-step methods that allows large variation in the integration step size and has unlimited stability regions (Hairer and Wanner, 1996; Freitas *et al.*, 2009; Sartori, 2014).

2.2. **Chemiosmotic gradients in the interface with early ocean and catalytic mineral setting**

For the temporal evolution of the concentration of the chemical species on the radiolytic fluid-ocean interface, we considered an isothermal diffusion model in continuous interface (see supplementary material for a detailed discussion). We considered the second Fick's law for the potential calculations. This provides the time and space diffusion effects. The diffusivity, $D_p$, is considered constant, given the high dilution of the species on the early ocean environment (Crank, 1979). Finally, we calculated the thermodynamic forces in gradients, such as electrochemical potentials and chemical potentials, in order to obtain the free energy at the interface.

The Electromotive force ($\phi$), which is associated to local redox potential, is calculated as discussed in the supplementary material. Besides, we calculated the local chemical potential $\mu_P$ from electromotive force $\phi$. The time range for diffusional migration, however, was arbitrarily chosen for a time around a week. Nevertheless, the difference of time range would change only the order of magnitude of the gradient region and does not change the values of the gradients' parameters.

We combined the thermodynamic parameters calculated with the energetic demand for the initial step of the acetyl-CoA synthesis pathway, used here as a reference for the protometabolic step, which reduces $CO_2$ to formaldehyde. In addition, we used the pH and local reducing potential to evaluate the plausibility of formation of structures composed of calcium carbonate ($CaCO_3$) with clusters of catalyst Fe-S minerals, like pyrite ($FeS_2$), mackinawite and greigite. For this analysis, we considered the radiolytic fluid-early ocean interface with temperature from 60ºC to 5ºC; and pressure around 0.25kbar, considered feasible for a natural radioactive environment (Dubessy *et al.*, 1988). Stability of Fe-S minerals is inferred from stability diagrams (potential/pH diagram).

3. **Results**

   3.1. **Kinetic model results for early ocean radiolysis**

Figure 1 shows the local chemical diversity of a radiolytic fluid in a natural radioactive environment as result of long-term radiolysis of mildly acidic (pH=5) early ocean. There is an initial growth of the concentration of products, characterizing the moment at which the radiolysis phenomenon dominates, providing the primary products. That lasts until the chemical reactions start to prevail, exhibiting a *quasi*-steady-state condition for the concentration of products. From the figure aforementioned, concentrations of $H_2$, $H^+$ and $OH^-$ ions, are several orders of magnitude higher than the other radical species, electrons ($e^-_{aq}$) and hydrogen peroxide ($H_2O_2$). This reflects the chemical instability of these species in respect to molecular ones. Comparing the difference in radionuclide concentration for the considered scenarios, the results show that this quantity affects the resulting chemical diversity. For example, Figure 1a shows results using radionuclide concentration from Witwatersrand region (correcting, as mentioned in method section, the activity of these to 4 Ga). On other hand, in Figure 1b, which represents a model based on radionuclides concentrations of chondrites, the $H_2$ reaches a concentration of up to three orders of magnitude lower than the case of Figure 1a (see also supplementary material for results in other scenarios).

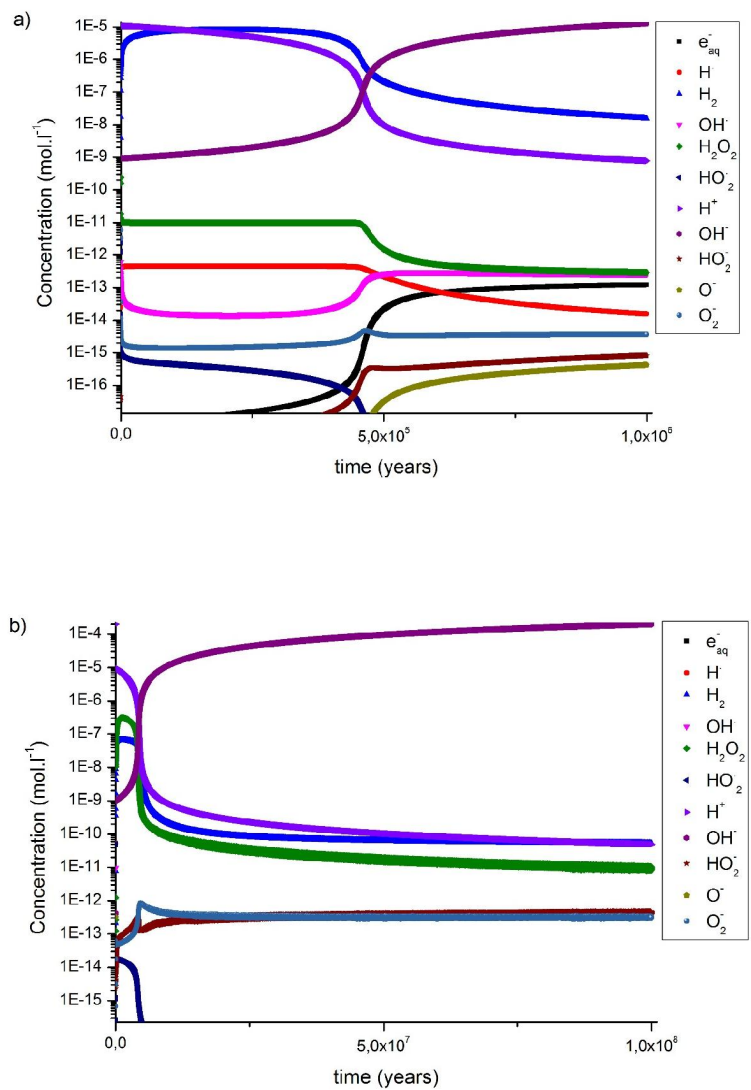

Figure 1- Results for long-term water radiolysis of early ocean production of chemical species. The concentrations are presented as a function of time in years. There are two sets of results considering: a) Concentration of radionuclides similar to Witwatersrand and a mildly acidic primitive ocean (pH = 5); and b) Concentration of radionuclides similar to Carbonaceous chondrite (see Table 1), which is used as control and considering a mildly acidic early ocean (pH = 5).

### 3.2. Chemiosmotic gradients formed in radiolytic fluid-early ocean interface

In Figure 2, we present the electromotive force (EMV,$\phi$) and chemical potential ($\mu$) distribution in the radiolytic fluid-early ocean interface. Its results of diffusion of prevailing products of radiolysis reported in last section to a mildly acid early ocean model. The figure shows that there is local free energy available from the chemical disequilibrium generated by water radiolysis to drive prebiotic transduction reactions; It is noteworthy, as shown in Figure 2, a local reducing environment whose electrochemical potential of a maximum of 230mV and chemical of a maximum 22kJ.mol$^{-1}$.

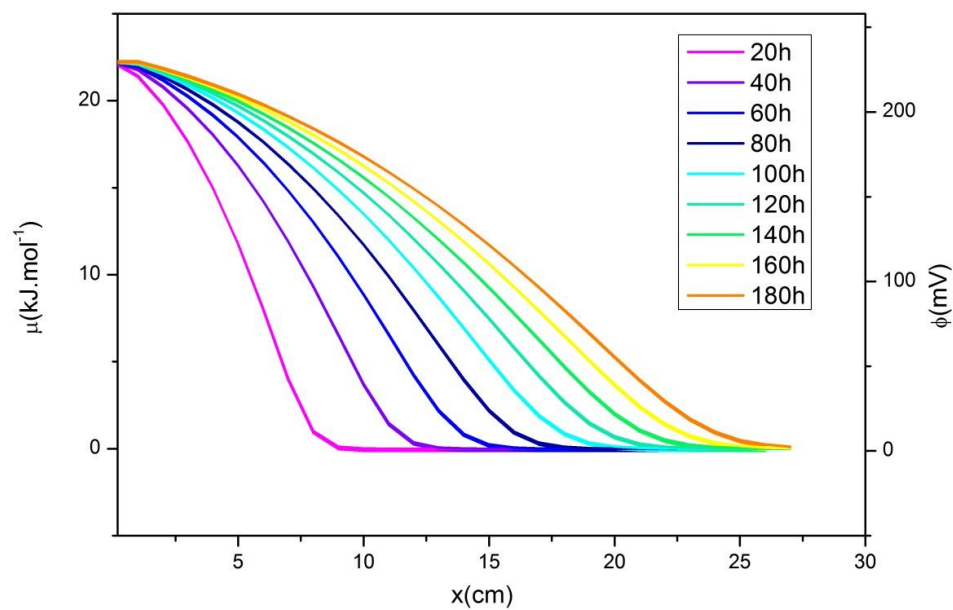

Figure 2 – Distribution of physicochemical parameters results in time and space from a one-dimensional diffusion model associated to radiolytic fluid-early ocean interface, for T = 5ºC. In x=0, it was considered the chemical diversity resulting from the numerical model for long-term mildly acidic (pH=5) early ocean water radiolysis. Each figure presents respectively: a) electromotive force ($\phi$); and b) chemical potential ($\mu$).

### 3.3. Analysis of precipitation of mineral catalysts in the interface

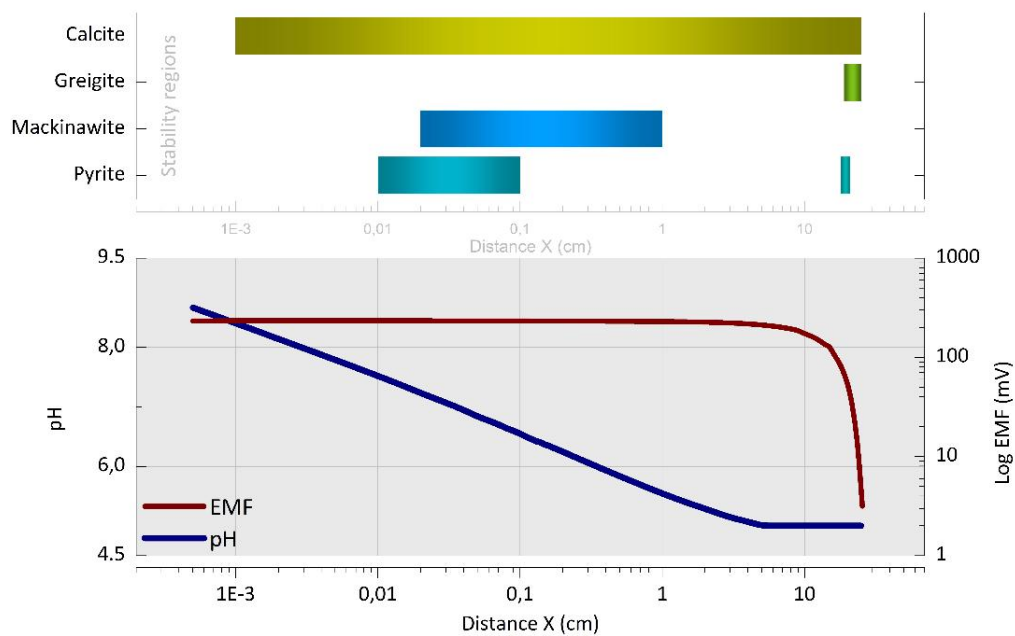

Figure 3 – Stability regions of carbonate, iron and sulphur minerals in the radiolytic fluid-early ocean interface. It presents the variation of pH and the log of the electromotive force (EMF) as a function of the distance from the radioactive source after stabilization of the values (after 180 hours of free-diffusion of the chemical species production).

At first approach, it is reasonable to expect that carbonates would precipitate in the radiolytic fluid-ocean interface as shown in Figure 3, because its precipitation is related to early ocean composition in contact to alkaline medium. However, due the typical temperature and pressure conditions associated to modern radioactive environments (Dubessy *et al.*, 1988; Savary and Pagel, 1997; Lin *et al.*, 2005a), it is expected the predominance of calcium carbonate in the form of calcite and the absence of aragonite in the radioactive environment, according to the stability diagram of these minerals (Garrels and Christ, 1965; Rickard *et al.*, 2007). Pyrite and calcite have already been reported at the Witwatersrand region, with calcite near the region of deposition of radioactive minerals (Hallbauer, 1986; Robb and Meyer, 1995; Omar *et al.*, 2003; Chivian *et al.*, 2008). Also, it is estimated the presence of catalytic minerals in regions near to early NRE (Ebisuzaki and Maruyama, 2017). In addition, the specific presence of mackinawite and greigite in the Witwatersrand environment, as the presence of pyrite in the region, is considered feasible (Hallbauer, 1986; Chivian *et al.*, 2008).

From the $\phi$ /pH diagram (Figure 3), it is observed that even in the presence of a low amount of sulphur and iron, some catalytic minerals are stable due to the strongly reducing conditions. The pyrite stability region is distributed in a limited region together with greigite and mackinawite close to the radioactive environment (less than 1 cm). Thus, Figure 3 illustrated numerically the possibility of formation of a mineral setting in NRE similar to those proposed in early AHV systems.

4. **Discussion**

The results of our kinetic model for long-term radiolysis compared to other models and geochemical measurements shows its adequacy to elucidate the physicochemical conditions of a habitable natural radioactive environment. As an example, the geochemical analysis of habitable NRE supports the resulted $H_2$-rich and reducing radiolytic fluid. Also, the final pH reached after around hundreds of thousands of years of radiolysis shows a alkaline medium, similar to the natural radioactive environment at which *Ca. D. audaxviator* was described (Chivian *et al.*, 2008).

Interestingly, the results showed also the occurrence of a *pH shift* due to long-term water radiolysis, which interchanges $OH^-$ and $H^+$ concentrations (see Figure 1). The inflection points on the $H^+$ and $OH^-$ curves occurs after $4.60 \times 10^5$ and $4.26 \times 10^6$ years considering, respectively, the radionuclide concentration from Witwatersrand and from chondrites. It shows the time when a natural radioactive environment as a modelled reaches an alkaline condition. The figure also indicates that in environments with higher concentrations of radionuclides, the shift occurs at earlier times, thus leading to different evolution paths for the geological settings. On other hand, in some models that consider an initial alkaline water medium, this pH shift occurs inversely, which means a lower final $H^+$ concentration (see Figure S1, S2 and S3).

Radiolysis produced chemicals with higher concentrations than others in radiolytic fluid as showed in Figure 1. In other words, after reaching a *quasi*-steady-state condition of the concentration, some products prevail in radiolytic fluid. Especially $OH^-$ and $H_2$ have influenced significantly the overall results for calculating the chemiosmotic gradients. The redox gradient, as example, associated to electromotive force (EMF; $\phi$) showed a maximum of 230mV, as presented in Figure 2. This is comparable to the one generated by the gradients in alkaline hydrothermal vents or on mitochondria of modern cells (around 200mV) (Barge *et al.*, 2012; Sojo *et al.*, 2016). Also, it is enough for driving the mentioned first step of Acetyl-CoA synthesis pathway. It is estimated that it occurs in 94mV, considering the pH gradient for the calculation (see figure S2).

The results shown in section 3.3 have pointed the feasibility of the precipitation of porous mineral interfaces that would contribute to the formation and maintenance of gradients and would contain catalytic Fe-S mineral clusters. The results support a hypothesis of transduction based on acetyl-CoA synthesis pathway in NRE. Considering the results in Figure 3, we proposed a model analogous for a possible primordial free energy conversion setting in a natural radioactive environment in early Earth in Figure 4. This Figure basically shows an analogy for the model of a requisite setting for emergence of life in AHV theory (see introduction).

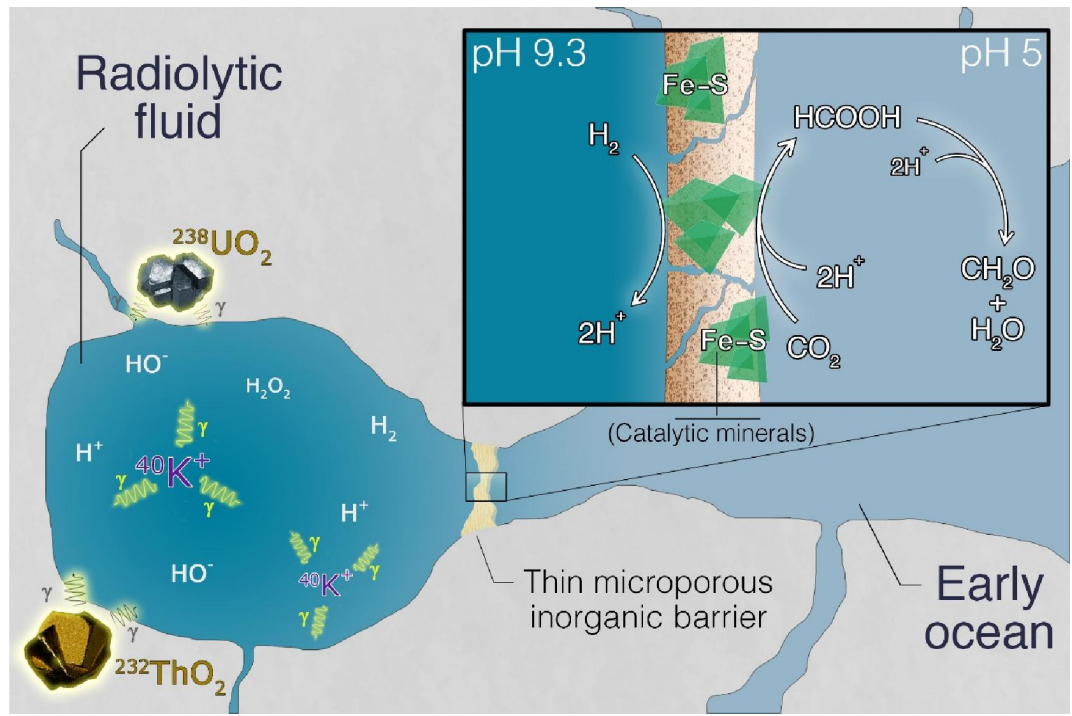

Figure 4 – Possible scenario for energy transduction on the interface between natural radioactive environment and Hadean ocean, based on the hypothesis for energy transduction in early alkaline hydrothermal system. It is considered Fe-S catalytic minerals that are stable on the interface (see section 3.3).

## 5. Conclusions

Geochemical analysis and metagenomics have shown that *Candidatus Desulforudis audaxviator* inhabits a reducing, $H_2$-rich, and alkaline environment (pH=9.3) (Chivian *et al.*, 2008). The local long-term water radiolysis is considered to have had an important contribution to this condition (Lin *et al.*, 2005b, 2006; Chivian *et al.*, 2008). However, those results did not specify how radiolysis contributed to this local condition or if it was specific for the local geological and mineralogical setting. It was a model specific related to gamma-ray radiolysis, on first moment due its high chemical diversity produced and due its effect in long range. However, the similarities mentioned above suggest that aspects of the model may provide perspectives about importance of this radiolysis to modern conditions of the habitable NRE such as those in Witwatersrand region.

In conclusion, we analysed NRE as possible scenarios for the emergence of life based on the similarity between physic-chemical condition of these environments and the conditions of alkaline hydrothermal vents. Our model showed that natural radioactive environment may have interesting conditions for application of AHV theory. Also showed that NRE in early Earth can represent novel sites for the precipitation of catalytic minerals, synthesis of prebiotic molecules and formation of chemical gradients necessary for the emergence of life. This indicates that, together with hydrothermal systems, they represent potential sites for the origin of life not only on Earth but also on other bodies of the Solar System, such as the icy moons of the giant planets, in the absence of sunlight or geothermal sources.

**Acknowledgements**

The authors thank FAPESP (projects 2016/06114-6 and 2016/08854-7), CAPES and CNPq (project 424367/2016-5) for the financial support and the Research Unit in Astrobiology (NAP/Astrobio – PRP/USP) for the institutional support. The authors also thank Serrapilheira Project number G-1709-20205.

**Disclosure Statement**

The authors declare no competing interests.

Lessons from Systems Biology. *Nature Reviews Microbiology*. 9:452–466.

Supplementary material for

Natural radioactive environments as sources of local disequilibrium for the emergence of life

Thiago Altair[1,5*], Larissa M. Sartori[2], Fabio Rodrigues[3], Marcio G. B. de Avellar[4], Douglas Galante[1,5]

[1] Brazilian Synchrotron Light Laboratory (LNLS), Brazilian Center for Research in Energy and Materials (CNPEM). Av. Giuseppe Máximo Scolfaro, 10000, 13083-100, Campinas/SP, Brazil.

[2] Instituto de Matemática e Estatística, Universidade de São Paulo. Rua do Matão, 1010, 05508-090, São Paulo/SP, Brazil.

[3] Departamento de Química Fundamental Instituto de Química, Universidade de São Paulo. Av. Prof. Lineu Prestes, 748, 05508-000, São Paulo/SP, Brazil.

[4] Instituto de Astronomia, Geofísica e Ciências Atmosféricas, Universidade de São Paulo. Rua do Matão, 1226, 05508-090, São Paulo/SP, Brazil.

[5] Instituto de Física de São Carlos, Universidade de São Paulo. Av. Trabalhador São-carlense, 400, 13566-590, São Carlos/SP, Brazil.

## Origins of life in hydrothermal systems and analogy with natural radioactive environments

Alkaline hydrothermal vents have less aggressive conditions for the emergence and maintenance of biological systems compared to other types of hydrothermal vents (such as black smokers), mainly because they have no direct contact with a magma chamber. Its local chemical disequilibrium is caused by the serpentinization chemical reaction (Lowell, 2002; Russell and Arndt, 2004; Martin *et al.*, 2008) as represented in equation (S1). It is proposed a free energy conversion condition for the vents in Lost City based on chemiosmotic gradients between alkaline hot hydrothermal fluid and a cold, mildly acidic ocean, capable of generating electromotive force (Martin *et al.*, 2008; Lane *et al.*, 2010; Branscomb and Russell, 2013; Sojo *et al.*, 2016; Lane, 2017). This interface is mediated by inorganic material, such as carbonate precipitates, making possible a thermal, pH and electrochemical gradient (Martin *et al.*, 2008; Lane *et al.*, 2010; Branscomb and Russell, 2013; Sojo *et al.*, 2016). Similar conditions of temperature and pH, and the predominance of reducing species are present in natural radioactive environments at where *Ca. D. audaxviator* was found (Lin *et al.*, 2005; Chivian *et al.*, 2008). Thus, one can evaluate the similarity of these radioactive environments with hydrothermal alkaline vents and even evaluate the possibility of associating energy transfer models to protobiological systems, as it was done for Lost City.

$$(Mg, Fe)_2 SiO_4 + H_2O + C \rightarrow Mg_3 SiO_5 (OH)_4 + Mg(OH)_2 + Fe_3 O_4 + H_2 + CH_4 + C_2 - C_5 \tag{S5}$$

## Energy transduction based in chemiosmosis in geologic systems in early Earth

Modern living systems make use of numerous mechanisms, mainly involving complex nanoengines driven by chemiosmotic gradients that work together with the metabolic pathways (Martin *et al.*, 2008; Lane and Martin, 2012; Branscomb and Russell, 2013; Russell *et al.*, 2013; Sojo *et al.*, 2016, 2017), as the example of the machinery involving ATP synthase (Boyer, 1997). In these mechanisms, the chemiosmotic gradients are maintained by semipermeable membranes that restrict the diffusion of chemical species creating a natural electrochemical potential, while the crossing through these membranes is mediated by specific and also complex converters, leading to the transduction of the chemical energy stored in the gradients (Branscomb and Russell, 2013). For the emergence of life in alkaline hydrothermal vents, such as those in Lost City Hydrothermal Fields, some authors have proposed an analogy with the modern biological electrochemical-driven mechanism, though replacing complex enzyme machinery with inorganic structures. These catalytic converters of free energy are driven by pH, electrochemical and thermal gradients, and are mainly based on mineral catalysts in hydrothermal vents environments (Nakamura *et al.*, 2010; Branscomb and Russell, 2013; Ang *et al.*, 2015).

The heated, reduced, alkaline hydrothermal fluid, along with the cold and mildly acidic oceanic solution, percolates micropores in the carbonate precipitate that is formed (Kelley, 2005). Precipitation of porous material in alkaline hydrothermal vents results in inorganic barriers - composed mainly of aragonite, $CaCO_3$ (Kelley, 2005) – with thickness of the order of micrometers (Kelley, 2005; Sojo *et al.*, 2016). The most important point of the model is the mechanism of catalysis by minerals composed of iron and sulfur – Fe-S minerals – or iron oxyhydroxide minerals, the called "green rusts"(Géhin *et al.*, 2002; Duval *et al.*, 2019). This Fe-S minerals, such as greigite ($Fe_3S_4$) or mackinawite (FeS), are speculated to have been incorporated into the precipitates of alkaline hydrothermal vents under the conditions of the early Earth. Considering early ocean and hydrothermal fluid conditions (Wächtershäuser, 1997; Sojo *et al.*, 2016), Fe-S clusters are estimated with structures such as those from minerals cited, being morphologically similar to important enzyme cofactors required for carbon fixation processes in the Acetyl-CoA synthesis pathway (Wood-Ljunghdahl)(Burcar *et al.*, 2015; Sojo *et al.*, 2016). The Acetyl-CoA pathway depends on a proton gradient for $CO_2$ reduction by $H_2$ using iron-sulfur proteins, such as ferredoxin (Nitschke and Russell, 2013; Herschy *et al.*, 2014; Sojo *et al.*, 2016). Notwithstanding, this pathway is an important one for the maintenance and evolution of autotrophic cells, not only to fix carbon, but also because of its capacity for the free energy conversion without ATP synthase. In addition, it is a very simple pathway, which, in few steps, converts $H_2$ and $CO_2$ to acetyl-CoA. This pathway exists in bacteria and archaea, and it is expected to have been present in the Last Universal Common Ancestor (LUCA)(Lane *et al.*, 2010). In summary, this carbonate structure incorporating Fe-S minerals, creating a microporous system of this material, makes the hydrothermal system a feasible site for $CO_2$ reduction pathway in a protometabolic analogous to the synthesis of acetyl- CoA. Its initial step, at which four electrons are transferred to carbon dioxide, is shown in equations (S2) and (S3).

$$H_2 \rightarrow 2H^+ + 2e^- \tag{S6}$$

$$CO_2 + 4H^+ + 4e^- \rightarrow CH_2O + H_2O \tag{S7}$$

**Physicochemical conditions of the natural radioactive environments and effects of water radiolysis in local chemical diversity**

Table S1 - Reactions involving products of radiolysis and their respective reaction rate constants

| | Chemical reaction | | | |
|---|---|---|---|---|
| | | **Reversible reactions** | | **pKa** |
| 1 | $H_2O$ | ↔ | $H^+ + OH^-$ | 13.999 |
| 2 | $H_2O_2$ | ↔ | $H^+ + HO_2^-$ | 11.65 |
| 3 | $OH$ | ↔ | $O^- + H^+$ | 11.9 |
| 4 | $HO_2$ | $\leftrightarrow$ | $O_2^- + H^+$ | 4.57 |
| 5 | $H$ | $\leftrightarrow$ | $e_{aq}^- + H^+$ | 9.77 |
| | | | | **Reaction rate (M/s)** |
| 7 | $H^+ + OH^-$ | $\rightarrow$ | $H_2O$ | $1.4 \times 10^{11}$ |
| 8 | $H_2O$ | $\rightarrow$ | $H^+ + OH^-$ | $K_7 \times K_2/[H_2O]$ |
| 9 | $H_2O_2$ | $\rightarrow$ | $H^+ + HO_2^-$ | $k_{10} \times K_3$ |
| 10 | $H^+ + HO_2^-$ | $\rightarrow$ | $H_2O_2$ | $5.0 \times 10^{10}$ |
| 11 | $H_2O_2 + OH^-$ | $\rightarrow$ | $H_2O + HO_2^-$ | $1.3 \times 10^{10}$ |
| 12 | $HO_2^- + H_2O$ | $\rightarrow$ | $H_2O_2 + OH^-$ | $k_{11} \times K_2/K_3 \times [H_2O]$ |
| 13 | $e_{aq}^- + H_2O$ | $\rightarrow$ | $H + OH^-$ | $1.9 \times 10^1$ |
| 14 | $H + OH^-$ | $\rightarrow$ | $e_{aq}^- + H_2O$ | $2.2 \times 10^7$ |
| 15 | $H$ | $\rightarrow$ | $e_{aq}^- + H^+$ | $k_{16} \times K_6$ |
| 16 | $e_{aq}^- + H^+$ | $\rightarrow$ | $H$ | $2.3 \times 10^{10}$ |
| 17 | $OH + OH^-$ | $\rightarrow$ | $O^- + H_2O$ | $1.3 \times 10^{10}$ |
| 18 | $O^- + H_2O$ | $\rightarrow$ | $OH + OH^-$ | $k_{17} \times K_2/K_4 \times [H_2O]$ |
| 19 | $OH$ | $\rightarrow$ | $O^- + H^+$ | $K_{20} \times K_4$ |
| 20 | $O^- + H^+$ | $\rightarrow$ | $OH$ | $1.0 \times 10^{11}$ |
| 21 | $HO_2$ | $\rightarrow$ | $O_2^- + H^+$ | $K_{22} \times K_5$ |
| 22 | $O_2^- + H^+$ | $\rightarrow$ | $HO_2$ | $5.0 \times 10^{10}$ |
| 23 | $HO_2 + OH^-$ | $\rightarrow$ | $O_2^- + H_2O$ | $5.0 \times 10^{10}$ |
| 24 | $O_2^- + H_2O$ | $\rightarrow$ | $HO_2 + OH^-$ | $k_{23} \times K_2/K_5 \times [H_2O]$ |
| 25 | $e_{aq}^- + OH$ | $\rightarrow$ | $OH^-$ | $3.0 \times 10^{10}$ |
| 26 | $e_{aq}^- + H_2O_2$ | $\rightarrow$ | $OH + OH^-$ | $1.1 \times 10^{10}$ |
| 27 | $e_{aq}^- + O_2^- + H_2O$ | $\rightarrow$ | $HO_2^- + OH^-$ | $1.3 \times 10^{10}/[H_2O]$ |
| 28 | $e_{aq}^- + HO_2$ | $\rightarrow$ | $HO_2^-$ | $2.0 \times 10^{10}$ |
| 29 | $2e_{aq}^- + 2H_2O$ | $\rightarrow$ | $H_2 + 2OH^-$ | $5.5 \times 10^{9}/[H_2O]$ |
| 30 | $e_{aq}^- + H + H_2O$ | $\rightarrow$ | $H_2 + OH^-$ | $2.5 \times 10^{10}/[H_2O]$ |
| 31 | $e_{aq}^- + HO_2^-$ | $\rightarrow$ | $HO_2^- + OH^-$ | $3.5 \times 10^9$ |
| 32 | $e_{aq}^- + O^- + H_2O$ | $\rightarrow$ | $OH^- + OH^-$ | $2.2 \times 10^{10}/[H_2O]$ |
| 33 | $H + H_2O$ | $\rightarrow$ | $H_2 + OH$ | $1.1 \times 10^1$ |
| 34 | $H + O^-$ | $\rightarrow$ | $OH^-$ | $1.0 \times 10^{10}$ |
| 35 | $H + HO_2^-$ | $\rightarrow$ | $OH + OH^-$ | $9.0 \times 10^7$ |
| 36 | $H + H$ | $\rightarrow$ | $H_2$ | $7.8 \times 10^9$ |
| 37 | $H + OH$ | $\rightarrow$ | $H_2O$ | $7.0 \times 10^9$ |
| 38 | $H + H_2O_2$ | $\rightarrow$ | $OH + H_2O$ | $9.0 \times 10^7$ |
| 39 | $H + HO_2$ | $\rightarrow$ | $H_2O_2$ | $1.8 \times 10^{10}$ |
| 40 | $H + O_2^-$ | $\rightarrow$ | $HO_2^-$ | $1.8 \times 10^{10}$ |
| 41 | $OH + OH$ | $\rightarrow$ | $H_2O_2$ | $3.6 \times 10^9$ |

| # | Reaction | | Products | Rate |
|---|---|---|---|---|
| 42 | $OH + HO_2$ | → | $H_2O + O_2$ | $6.0 \times 10^9$ |
| 43 | $OH + O_2^-$ | → | $OH^- + O_2$ | $8.2 \times 10^9$ |
| 44 | $OH + H_2$ | → | $H + H_2O$ | $4.3 \times 10^7$ |
| 45 | $OH + H_2O_2$ | → | $HO_2 + H_2O$ | $2.7 \times 10^7$ |
| 46 | $OH + O^-$ | → | $HO_2^-$ | $2.5 \times 10^{10}$ |
| 47 | $OH + HO_2^-$ | → | $HO_2 + OH^-$ | $7.5 \times 10^9$ |
| 48 | $HO_2 + O_2^-$ | → | $HO_2^- + O_2$ | $8.0 \times 10^7$ |
| 49 | $HO_2 + HO_2$ | → | $H_2O_2 + O_2$ | $7.0 \times 10^5$ |
| 50 | $HO_2 + O^-$ | → | $O_2 + OH^-$ | $6.0 \times 10^9$ |
| 51 | $HO_2 + H_2O_2$ | → | $OH + O_2 + H_2O$ | $5.0 \times 10^{-1}$ |
| 52 | $HO_2 + HO_2^-$ | → | $OH + O_2 + OH^-$ | $5.0 \times 10^{-1}$ |
| 53 | $2O_2^- + 2H_2O$ | → | $H_2O_2 + O_2 + 2OH^-$ | $1.0 \times 10^2/2[H_2O]$ |
| 54 | $O_2^- + O^- + H_2O$ | → | $O_2 + 2OH^-$ | $6.0 \times 10^8/[H_2O]$ |
| 55 | $O_2^- + H_2O$ | → | $OH + O_2 + OH^-$ | $1.3 \times 10^{-1}$ |
| 56 | $O_2^- + HO_2^-$ | → | $O^- + O_2 + OH^-$ | $1.3 \times 10^{-1}$ |
| 57 | $2O^- + H_2O$ | → | $HO_2^- + OH^-$ | $1.0 \times 10^9/[H_2O]$ |
| 58 | $O^- + H_2$ | → | $H + OH^-$ | $8.0 \times 10^7$ |
| 59 | $O^- + H_2O_2$ | → | $H_2O$ | $5.0 \times 10^8$ |
| 60 | $O^- + HO_2^-$ | → | $O_2^- + OH^-$ | $4.0 \times 10^8$ |

**Free Energy and establishment of chemiosmotic gradients in the interface with early ocean**

For the temporal evolution on the concentration of the chemical species on the radiolytic fluid-ocean interface, we considered an isothermal diffusion model in continuous interface. It was considered the second Fick's law, as presented in equation (S9), since radiolysis results in very low concentration of products. This provides the time and space diffusion effects. The diffusivity, $D_p$, was considered constant, given the high dilution of the species on the primitive ocean environment (Crank, 1979).

$$J_{N,K} = -D_P \frac{\partial C_K}{\partial x} \tag{S8}$$

$$\frac{\partial C}{\partial t} = D_K \cdot \frac{\partial^2 C}{\partial x^2} \tag{S9}$$

Equation (S9) relates the variation of $C_K$, the concentration of the chemical specie $K$ in mol.l$^{-1}$ to the space $x$ and the time $t$, given $D_K$ (the diffusion coefficient of product $K$ in cm$^2$.s$^{-1}$, considered for a temperature of 5ºC, a value within the temperature range that predominates in the early oceans (Pinti, 2005), just as in modern oceans (Sverdrup, H.; Johnson, M.; Fleming, 1970)). To find the solution for equation (S5) we used the *NSolve* package from *Wolfram Mathematica* assuming that in *x = 0* the concentrations were the final concentration calculated in our kinetic model and initial oceanic pH is 5.0 - to have a comparative to the models of hydrothermal vents (Russell and Arndt, 2004; Sojo *et al.*, 2016) - or 7.0, to have a control on pH effects on water radiolysis.

For chemiosmotic gradients and free energy calculations in the radiolytic fluid-ocean interface, thermodynamic forces in gradients needed to be calculated, such as electrochemical potentials (or electromotive force, E.M.F.), and chemical potentials ($\mu_P$). E.M.F. is calculated using the Nernst equation (S6, below(S10)) considering different ion activities inside and outside the system (the ocean).

$$\phi_{ion} = \frac{RT}{zF} ln \frac{C_{ion,ocean}}{C_{ion,system}} \qquad (S10)$$

In equation S9, $R$ is the universal gas constant of 8.314 J (mol.K)$^{-1}$; $T$ is the temperature of the solution in K; $z$ the ion electrical charge; $C$ the ion concentration and $F$ being the Faraday constant, equivalent to 96.485 C.mol$^{-1}$.

Besides E.M.F., as a parameter of major importance, $\mu_P$ is related to the E.M.F., as shown in equation (S7).

$$\Delta\mu_{ion} = Fz\phi_{ion} \qquad (S11)$$

**Free energy transfer mechanism between geological and proto-biological systems in natural radioactive environments**

The basis for this transduction mechanism is assumed to be a protometabolic chemiosmotic gradient-driven pathway analogous to acetyl-CoA synthesis (Martin *et al.*, 2008; Russell *et al.*, 2013; Herschy *et al.*, 2014; Sojo *et al.*, 2016; Branscomb *et al.*, 2017). In this analogous protometabolic pathway, there is no presence of any enzyme and transduction occurs necessarily through some mineralogical structure. Thus, we evaluated the plausibility of the formation of microporous structures composed of calcium carbonate ($CaCO_3$) with mineral catalysts' clusters, such as Fe-S minerals, similarly to the case of hydrothermal vents. For this analysis, we considered pH conditions, local redox potential and temperature (considering the radiolytic fluid-early ocean interface with temperature from 60ºC to 5ºC) and pressure around 0.25 kbar. Stability of sulfur minerals is inferred from the stability diagram (potential/pH diagram) of mackinawite and greigite in water at 25ºC considering (dissolved) sulfur concentration of $10^{-6}$ M and iron concentration of $10^{-3}$ M(Rickard *et al.*, 2007); stability of carbonate is based on pH gradient and pyrite is based on stability diagram considering $10^{-6}$M of total sulfur dissolved(Garrels and Christ, 1965).

**Distribution of $H_2$ concentration, pH and electrochemical potential in the radiolytic fluid-early ocean interface**

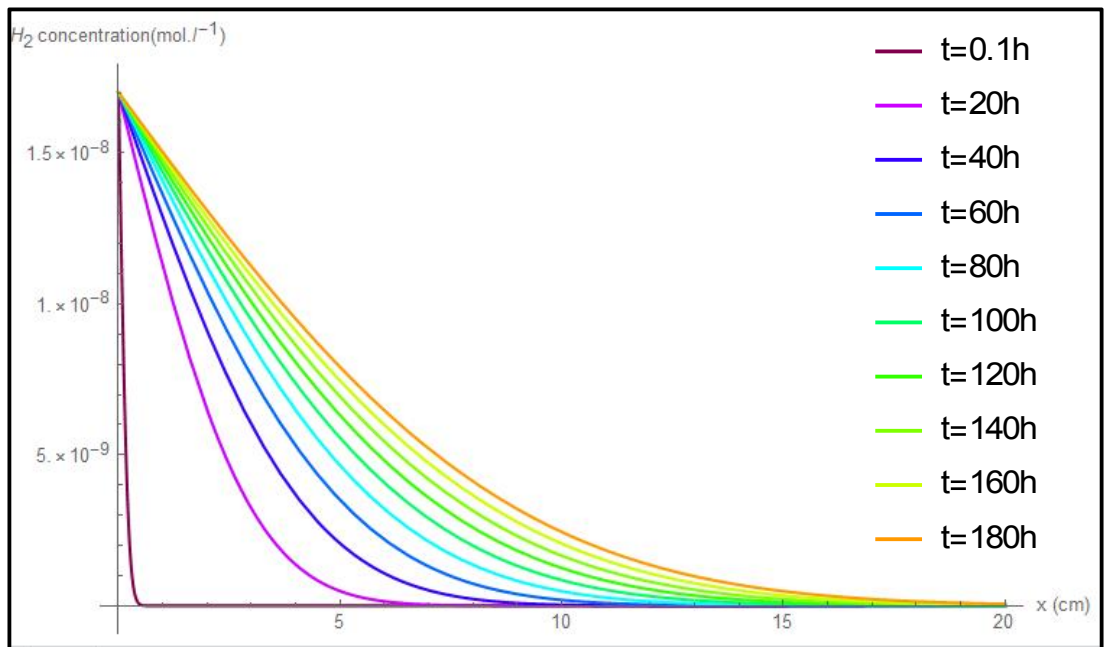

Figure S5 - Distribution of $H_2$ in time and space from one-dimensional diffusion model based on Fick's law in radiolytic fluid-early ocean interface, considering the constant diffusion coefficient and aqueous medium at T = 5 ° C. In this model, at x = 0 it is considered the final $H_2$ concentration in modelled natural radioactive environment, after ca. t = $7.5 \times 10^5$ years, as calculated based on the kinetic model presented on Figure 1.

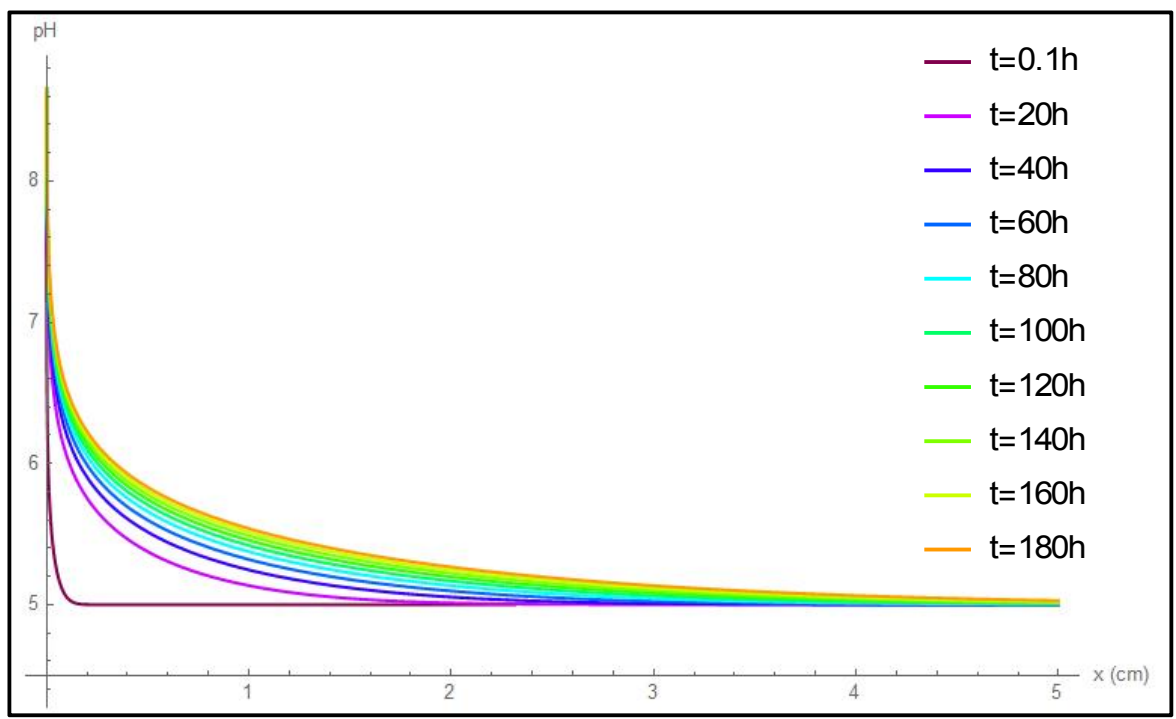

Figure S6 - pH gradient as a function of time and space at the interface region between a radioactive environment at x = 0 and the mildly acid early ocean, considering pH = 5 and temperature T = 5 ° C. For this calculation, it was considered a one-dimensional diffusion model for $H^+$ and $OH^-$ ions by Fick's law and considering constant diffusivity. At *x = 0*, it is considered final pH of a radioactive environment as calculated based on the kinetic model presented in Figure 1.

*Kinetic model results for several scenarios*

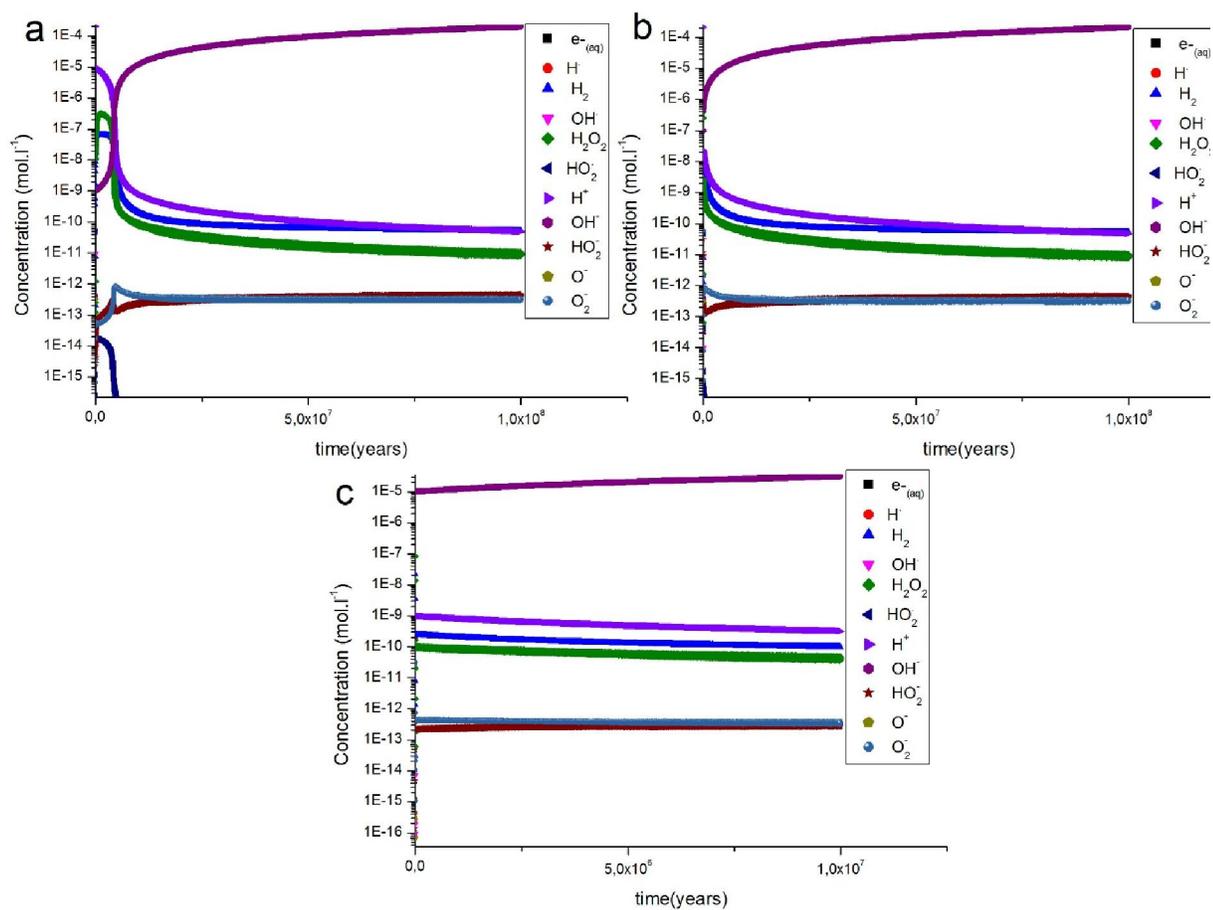

Figure S7 - Kinetic model results for water radiolysis in early natural radioactive environments using several scenarios with Chondrite typical radionuclide concentration considering three initial pH: a) pH=5; b) pH=7 and c) pH=9.

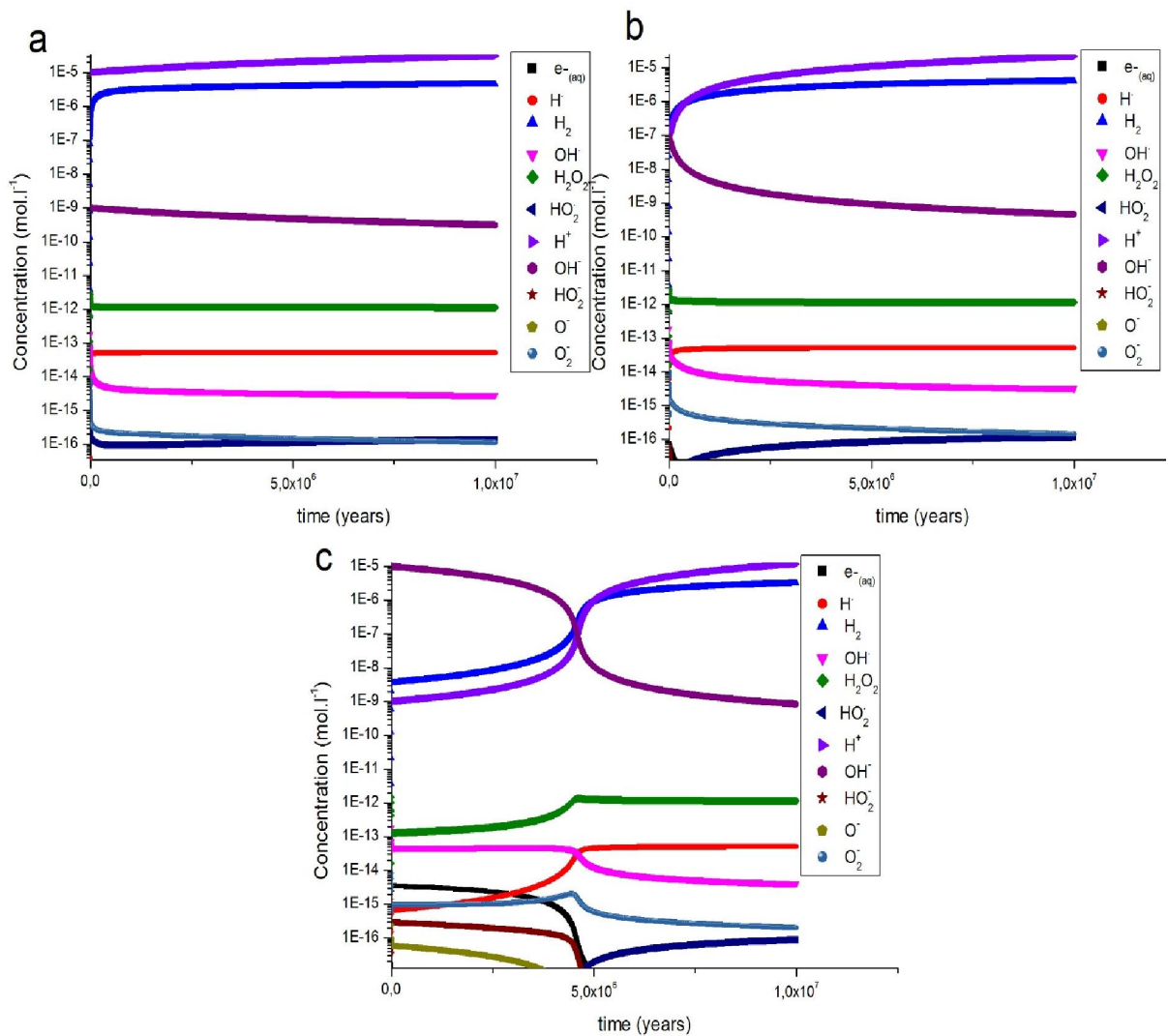

Figure S8 - Kinetic model results for water radiolysis in early natural radioactive environments using several scenarios with radionuclide concentration related to Witwatersrand non-mineralized strata considering three initial pH: a) pH=5; b) pH=7 and c) pH=9.

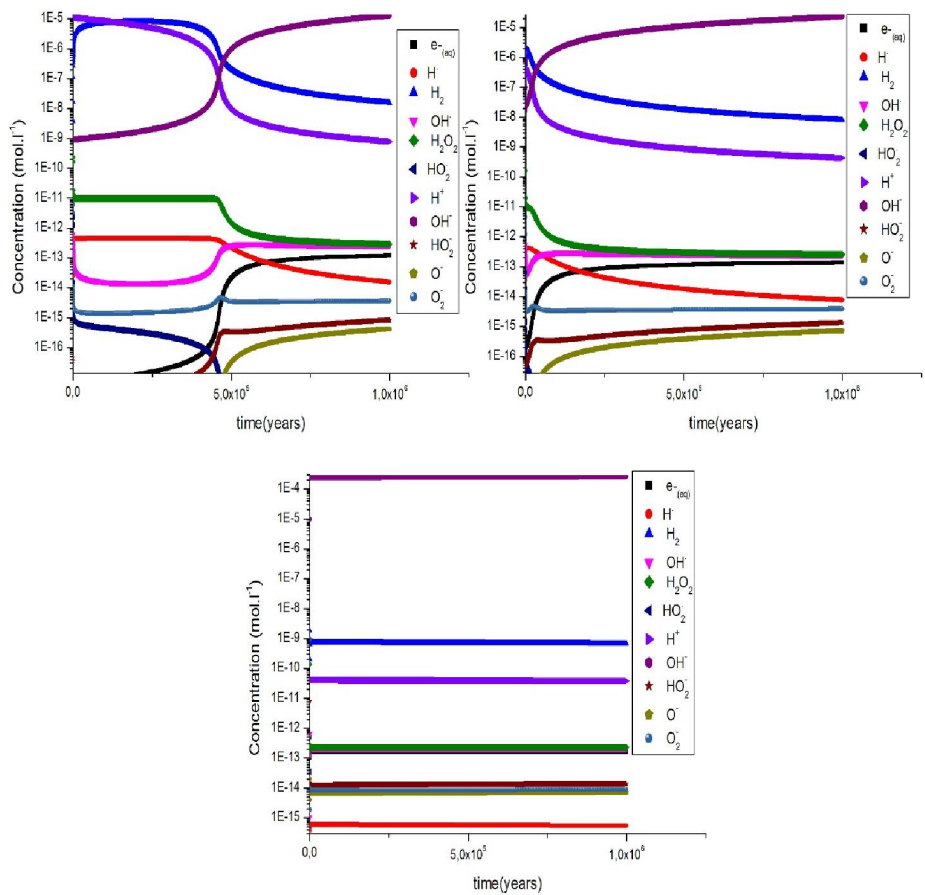

Figure S9 - Kinetic model results for water radiolysis in early natural radioactive environments using several scenarios with radionuclide concentration related to Witwatersrand non-mineralized strata considering three initial pH: a) pH=5; b) pH=7 and c) pH=9.